\documentclass[sigconf]{acmart}
\usepackage{graphicx} 
\usepackage{hyperref}
\usepackage{csquotes}
\usepackage{enumitem}
\usepackage{tikz}
\usetikzlibrary{shapes.geometric, arrows, arrows.meta, calc, fit}
\usepackage{enumitem}
\usepackage{colortbl}
\usepackage{xcolor}
\usepackage{multirow}
\usepackage{booktabs}
\usepackage{balance}

\title{The Case for Vibe Modeling: A Missing Step in AI-Based Trustworthy Software Development}
\renewcommand{\shorttitle}{The Case for Vibe Modeling}

\author{Shalini Chakraborty}
\affiliation{%
  \institution{University of Bayreuth}
  \city{Bayreuth}
  \country{Germany}
}
\email{shalini.chakraborty@uni-bayreuth.de}
\orcid{0000-0002-9466-3766}

\author{Michael Mittermaier}
\affiliation{%
  \institution{University of Regensburg}
  \city{Regensburg}
  \country{Germany}
}
\email{michael.mittermaier@ur.de}
\orcid{0000-0003-4235-7414}

\author{Judith Michael}
\affiliation{%
  \institution{University of Regensburg}
  \city{Regensburg}
  \country{Germany}
}
\email{judith.michael@ur.de}
\orcid{0000-0002-4999-2544}

\date{March 2026}
\acmConference[XXX]{XXX}{2026}{XXX}
\begin{document}

\begin{abstract}
Large Language Models (LLMs) are increasingly used to generate software artifacts from natural language prompts. While this enables rapid prototyping and lowers the barrier to software creation, it also introduces challenges related to understanding, validation, traceability, and trust. In this paper, we argue that current AI-based development practices focus too heavily on the direct generation of code and insufficiently on intermediate representations that preserve human intent and support reasoning about system behavior. We argue for \emph{vibe modeling} as a lightweight intermediate abstraction between natural language interaction and code generation. To explore its potential, we present a student survey study that examines perceptions of LLM output understanding, validation effort, trust and the perceived usefulness of vibe modeling across several AI-assisted development scenarios. Our results are intended to inform future studies for trustworthy and explainable AI-based software engineering via vibe modeling.
\end{abstract}


\maketitle

\section{Introduction}
Large Language Models (LLMs) are increasingly used to generate code and other software artifacts from natural-language prompts. This shift has made software creation faster and more accessible, and it has led to new interaction styles often described as \emph{vibe coding}~\cite{sarkar2025vibe}. However, direct generation from prompts also introduces a serious challenge: \textbf{developers must decide whether they can trust the artifacts they did not create and cannot
easily justify or reproduce.}

In software engineering (SE), trust is not merely a vague positive attitude toward a tool. It is closely related to a developer's ability to understand, validate, and control the resulting artifact. Yet much of the current literature and many practical tools still treat trust implicitly, without a clear operational account of what supports it or undermines it. Prior work suggests that trustworthiness in AI systems depends strongly on transparency, explainability, and traceability~\cite{mora2021traceability,baltes2026need}. In AI-assisted software development, however, the generation process often remains opaque: LLM-based tools can produce plausible outputs while exposing little of the reasoning, assumptions, or constraints that led to them~\cite{chazette2020explainability}. This creates a validation burden for developers, who must inspect not only the final code but also the hidden decisions embedded in its construction~\cite{shah2026explainability}.
Explainability and traceability are therefore highly relevant to AI-assisted development, but most existing approaches focus on explaining the \emph{output} of an AI system rather than supporting the \emph{process} by which software artifacts are produced. 
We argue that this gap motivates \emph{vibe modeling}~\cite{cabot2025vibe}: a lightweight intermediate abstraction that captures intent, constraints, assumptions, and rationale during AI-assisted development. Vibe modeling is not intended to replace formal modeling approaches. Rather, it preserves the exploratory and conversational character of LLM-based development while introducing a structured artifact that can be reviewed, versioned, and traced. In this sense, vibe modeling may help bridge the space between informal user intent and executable code.

To explore this idea, we report an initial empirical study with students who evaluated LLM-assisted development scenarios with and without intermediate modeling support. We articulate three claims that motivate the study: First, vibe modeling may improve the \textbf{understandability} of AI-generated artifacts by making intermediate decisions explicit. Second, it may reduce \textbf{validation effort} by shifting part of the reasoning burden from opaque code to a more inspectable abstraction. Third, it may support \textbf{trust} by preserving links between evolving intent, model changes, and code changes. Overall, rather than claiming that vibe modeling solves the trust problem, we investigate whether it is perceived as a meaningful step toward more trustworthy AI-assisted software development. 
Before establishing evidence for the above, in this paper, we report students' perspectives on vibe modeling and their perceptions of trust in LLM-generated output. The study follows three RQs: \textbf{RQ1:} What concerns do students have about LLM-generated code?, \textbf{RQ2:} Which factors do students identify as increasing and decreasing trust in the presented scenarios?, and \textbf{RQ3:} How do students perceive vibe modeling as an intermediate step between natural-language requirements and code generation?

\section{Related Work}
\subsection{Trust in AI-Assisted Software Development}
The increasing adoption of LLMs for software development has raised important questions regarding developers' trust in AI-generated artifacts. Empirical studies have shown that developers often struggle to assess whether generated artifacts satisfy intended requirements, particularly when the underlying reasoning process remains opaque \cite{baltes2026need,shah2026explainability}.
Research on trust in automation suggests that trust is closely linked to a user's ability to understand and predict system behavior \cite{lee2004trust}. In SE contexts, this implies that developers require mechanisms that support validation, inspection, and justification of AI-generated outputs so they can reproduce them. Existing AI-assisted development tools largely focus on code generation itself, leaving open the question of how trust can be established during the generation and regeneration process rather than only after artifacts have been produced.

\subsection{Explainability, Transparency, and Traceability}
Explainable AI (XAI) seeks to make system decisions understandable to human stakeholders and thereby support trust, accountability, and informed decision-making \cite{adadi2018peeking,arrieta2020explainable}. Explainable software engineering aims to make system behavior understandable to stakeholders and to support debugging, validation, and maintenance~\cite{chazette2022explainable}. Traceability complements this by linking requirements, models, and code across the development lifecycle~\cite{mora2021traceability}. Traceability has been shown to support maintenance, verification, compliance, and change impact analysis. However, traditional traceability approaches assume the existence of relatively stable development artifacts. 
In AI-assisted development, where software is increasingly generated through conversational interactions with LLMs, important design decisions may remain undocumented and difficult to reconstruct. Consequently, explainability and traceability remain necessary but insufficient for capturing the evolving intent behind AI-generated artifacts.

\subsection{Model-Based Software Engineering as an Intermediate Abstraction}

Model-Based Software Engineering (MBSE) promotes the use of abstract representations to support communication, analysis, and automation throughout software development \cite{schmidt2006model,chakraborty2024modelling}.
Numerous studies have demonstrated the benefits of modeling for improving system understanding, facilitating stakeholder communication, and supporting automated generation processes~\cite{brambilla2017model}. Nevertheless, traditional modeling approaches often require substantial expertise and upfront formalization effort~\cite{chakraborty2023we}. Such characteristics can conflict with the exploratory and iterative nature of AI-assisted development, where users frequently refine requirements through conversational interactions with generative systems.
Recent discussions around AI-supported modeling and model generation suggest that abstractions remain valuable even when software artifacts can be generated automatically \cite{sarker2022ai}. However, there remains a gap between informal natural-language prompts and formal software models. This gap motivates the exploration of lightweight intermediate abstractions that preserve intent while remaining accessible to developers.

\subsection{Vibe Coding and Emerging AI-Centric Development Practices}

Recent work has described a growing development style known as \emph{vibe coding}, in which users iteratively interact with LLMs to generate software artifacts through natural-language instructions rather than conventional programming activities \cite{sarkar2025vibe}. Vibe coding emphasizes rapid experimentation and accessibility, enabling users with varying technical backgrounds to create software systems.
While this paradigm lowers barriers to software creation, it also amplifies concerns regarding artifact ownership, accountability, and understanding. Generated code may evolve through multiple conversational iterations without leaving explicit records of assumptions, design rationale, or requirement changes. As a result, developers may find it difficult to explain how a system was produced or to justify confidence in its correctness.
The concept of \emph{vibe modeling} builds upon these observations. Rather than replacing existing modeling techniques, vibe modeling seeks to introduce a lightweight intermediate artifact that captures intent, rationale, and constraints throughout AI-assisted development. In doing so, it aims to bridge the gap between conversational software generation and established principles of explainability, traceability, and software quality assurance.

\section{Method}
We conducted an exploratory questionnaire study with students from two universities. We focus on students because they are actively learning vibe coding and modeling and can articulate when one or both are needed in specific development settings, making them well-positioned to reveal opportunities and challenges of introducing vibe modeling into practice.
The survey was designed around four development scenarios that vary in workflow and in the presence or absence of intermediate modeling support.
\subsection{Survey Instrument and Participation}

The questionnaire and analysis materials are available online~\cite{anon2026_vibemodeling}.
The survey was administered via a university-approved online platform. No personally identifying information (e.g., names, email addresses, student IDs, or IP addresses) was collected. Participants were recruited from two universities from their respective software engineering courses of the summer semester, and the survey remained open for one week.
Ethical approval was obtained prior to data collection. Participation was voluntary and participants could withdraw at any time without giving a reason and without consequences.
\subsection{Data Collection}
We first ran a pilot with 18 students to check clarity of the questionnaire, reasons of trust/no trust and, based on student feedback, modified the survey. The final survey was then open online for one week, yielding a total of 17 responses from both universities.

\subsection{Data Analysis}
Given the small, exploratory sample, we focus on descriptive analysis~\cite{ko2006linguistic}.
All closed-ended items were analyzed using basic descriptives (counts, percentages, means, and medians) to summarize students' perceptions of productivity, concerns of using LLM, and trustworthiness across the four scenarios.
For the trust-related items, we inspected endorsements of individual concern statements (e.g., lack of transparency, potential bias, difficulty of verification) to identify the most frequently reported issues. 
Open-ended responses were analyzed using a qualitative, coding-based approach. We first performed an initial reading of all comments, then inductively grouped them into categories capturing perceived benefits (e.g., structure, interpretability, validation support) and drawbacks (e.g., added effort, required skills, unclear value). These categories were refined iteratively and used to organize the findings in the results section. 

\subsection{Threats to Validity}


\subsubsection{External validity.} The study included 17 student participants from two universities. The small sample size and focus on students limit the generalizability of the findings to professional developers and other contexts. However, students are an appropriate population for this exploratory study, as they are actively learning software modeling and frequently use AI-assisted development tools.

\subsubsection{Construct validity.} Participants evaluated hypothetical scenarios rather than using an implemented vibe modeling environment, meaning the results reflect perceptions rather than observed behavior. To reduce this threat, the scenarios represented realistic AI-assisted development workflows, and the survey was refined through pilot studies by improving wording and providing a definition of vibe modeling.

\subsubsection{Internal validity.} Individual responses may have been influenced by participants' programming experience, AI familiarity, or attitudes toward automation. We collected demographic and experience-related information to support interpretation of these factors. In addition, survey questions were optional to avoid forcing uncertain or low-quality responses. Consequently, not all participants answered every question, resulting in some missing data.

\subsubsection{Reliability.} Qualitative coding is inherently interpretive and may introduce researcher bias. To improve reliability, codes were derived systematically from the data, and the study materials and coding artifacts will be made publicly available~\cite{anon2026_vibemodeling} to support transparency and reproducibility.
\section{Scenarios}
\begin{figure}
    \centering
    \includegraphics[width=\linewidth]{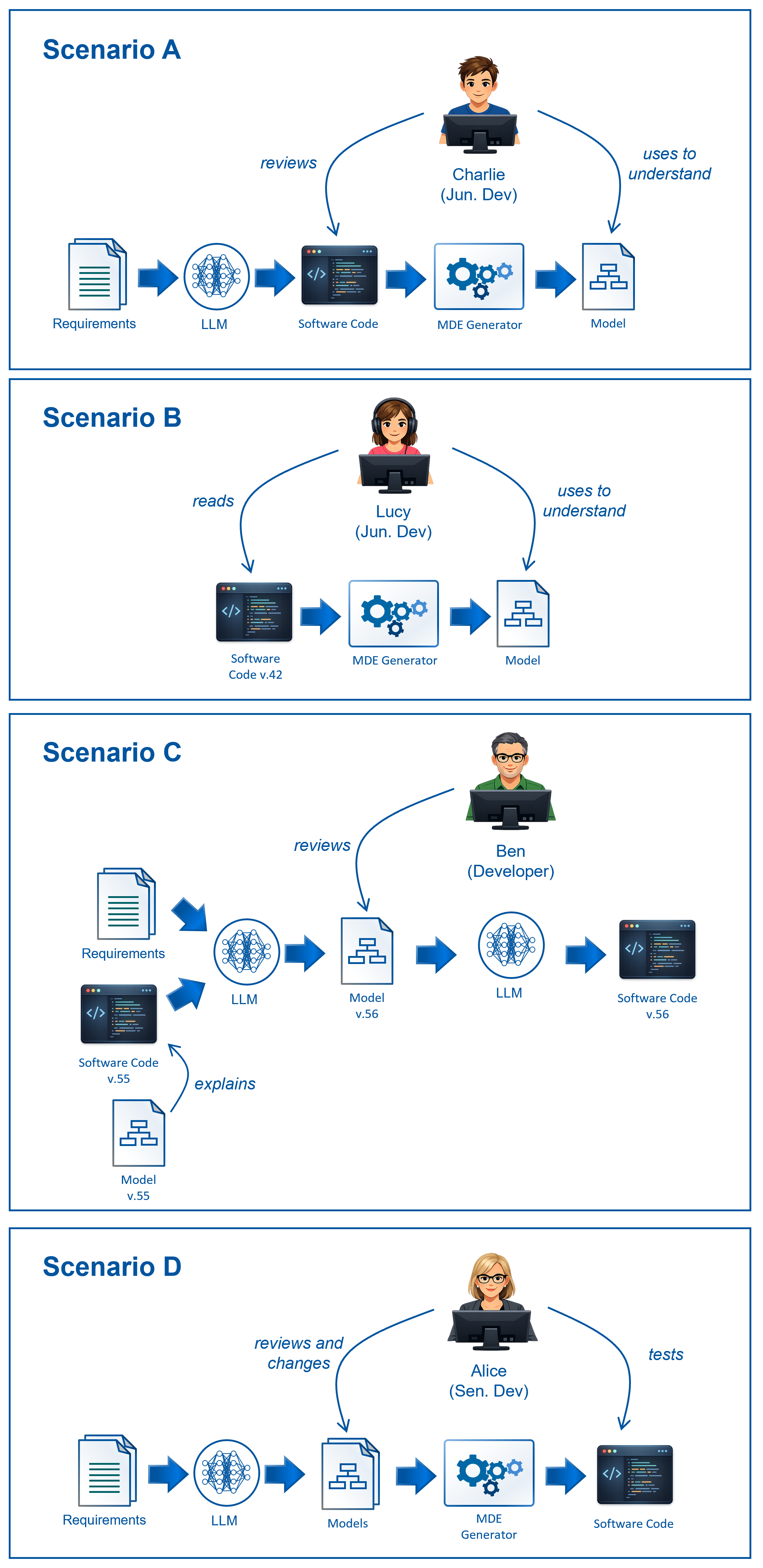}
    \caption{The four scenarios presented in the survey}
    \label{fig:scenarios}
\end{figure}

We use four scenarios to explore how students perceive AI-generated code, intermediate models, and vibe modeling (see~\autoref{fig:scenarios}). The scenarios cover direct code generation, brownfield change, model-first development, and code-to-model reconstruction.
In the scenarios, the term \emph{model} refers to a lightweight, human-readable representation such as a class diagram, component diagram, or architecture sketch that supports understanding and review.

These scenarios differ along two dimensions: the direction of abstraction flow and the development context. Scenarios A and B start from code and use models to support understanding, while scenarios C and D start from requirements or change requests and use models before code is modified or generated. Scenarios A and D represent greenfield or code-generation-centric settings, whereas B and C capture brownfield maintenance and evolution. Together, the scenarios let us compare how students perceive model reconstruction, model-mediated change, and model-first generation, and whether vibe modeling is seen as helpful for understanding, validation, and control.
Detailed descriptions of scenarios are added to the supplementary material~\cite{anon2026_vibemodeling}.
\section{Results}
\subsection{RQ1: Student Concerns about LLM-Generated Code}
We start the survey by asking students three Likert-scale statements about their
everyday experience with LLMs: \emph{LLMs will increase my productivity in software
development}, \emph{It is time consuming to validate LLM generated code}, and
\emph{It is difficult to accommodate changes into LLM generated code}.
On our 1--101 scale (1 = fully agree, 101 = fully disagree), most responses to all three items fall below the neutral midpoint, with several very low values, indicating that students generally agree that LLMs increase productivity but also see validation
effort and handling code changes as notable pain points. 
For productivity, 12 responses were below the neutral midpoint of 50.
This pattern indicates that most students tend to agree that LLMs increase their productivity.
However, for validation effort 11 values fall below 50 (with three values at below 5) and for change accommodation, 9 responses are below 50. 
This distribution suggesting that
students commonly experience validation of LLM-generated code as time consuming. A majority agrees that it is difficult to accommodate changes into LLM-generated code, but there is also a non-trivial minority who are neutral or disagree, suggesting that experiences with evolving such code vary more across students.

Across scenarios, respondents reported a broad range of concerns about over relying on AI-generated code as mentioned by a participant \enquote{\emph{Becoming over reliant and people not checking if the AI-generated code is valid}}.
For each scenario, we asked \emph{How concerned would you be about becoming overreliant on such an LLM in this work process?} (1 = very likely,
101 = not likely).
In Scenario~A, all 12 responses are below the neutral midpoint of 50, indicating clearly elevated concern about
overreliance in the code-first workflow.
Scenario~C shows a similar pattern (out of 9 responses, 7 values below 50)), suggesting that students are also
worried about overreliance when an LLM updates an existing system model
before code changes.
Scenario~B is more mixed (out of 10, 6 values below
and 4 at or above 50), while Scenario~D is closest to neutral
(Out of 11 responses, 4 below and 5 at or above 50), implying
that model-first development for a new system raises fewer overreliance
concerns than the more code-centric workflows.
Overall, these results suggest that students are concerned about
becoming overreliant on LLMs, particularly when the LLM is deeply
embedded in the development flow and less buffered by intermediate
models.

Students also highlighted the lack of transparency in the LLM’s reasoning process, the possibility of biased outputs, and the unpredictable variability in result quality, \enquote{\emph{In any conditions, issues with the final code are always unpredictable}}. The risk that important details or edge cases might be overlooked, that the LLM might lack sufficient knowledge of the specific project context, and that responsibility for mistakes cannot be assigned to the AI system are also mentioned. \enquote{\emph{Does an LLM have enough context on the non-technical aspects of projects to accurately model complex codebases and implementation decisions?}} 

\subsection{RQ2: Factors Increasing and Decreasing Trust}
To address RQ2, we examined which trust-related statements participants selected for each scenario (Trust-A, Trust-B, Trust-C, Trust-D) and mapped the corresponding item codes to the labels defined in the codebook.
This allows us to see not only whether students trusted the workflows overall, but also \emph{why} they considered specific AI-assisted outcomes more or less trustworthy in different settings. Table~\ref{tab:scenario-trust} and \ref{tab:scenario-notrust} shows details of all the reasons to trust and not trust the LLM output for each scenario.

\begin{table*}[t]
  \centering
  \caption{Scenario gist and most endorsed trust reasons (number of respondents).}
  \label{tab:scenario-trust}
  \begin{tabular}{p{0.06\textwidth}p{0.32\textwidth}p{0.54\textwidth}}
    \toprule
    Scenario & Gist & Most frequently selected trust reasons (count) \\ \midrule

    A &
    Code-first, then reconstruct a higher-level model from LLM-generated code to understand and verify behavior. &
    The LLM allows human oversight or correction (10);
    the LLM provides transparent reasoning or intermediate steps (9);
    the results can be verified independently (7);
    the results appear plausible and logically consistent (6);
    the explanations are understandable (6). \\[0.6em]

    B &
    Use model reconstruction on an existing long-running code base to understand the system before continuing development with LLM-generated code. &
    The results can be verified independently (7);
    the results align with my own expectations or expertise (6);
    the outputs are supported by data or evidence (6);
    the results appear plausible and logically consistent (5);
    the results are sufficiently detailed and precise (5). \\[0.6em]

    C &
    Update an existing system’s model (based on current code and a new requirement) with an LLM, review the model, then generate code changes. &
    The results appear plausible and logically consistent (6);
    the results are sufficiently detailed and precise (5);
    the LLM provides transparent reasoning or intermediate steps (5);
    the results align with my own expectations or expertise (5);
    the LLM allows human oversight or correction (5). \\[0.6em]

    D &
    Generate and review system models (architecture, data, behavior) with an LLM for a new system before deterministic code generation. &
    The results are sufficiently detailed and precise (7);
    the results appear plausible and logically consistent (5);
    the LLM provides transparent reasoning or intermediate steps (5);
    the outputs are supported by data or evidence (5);
    the results can be verified independently (5). \\

    \bottomrule
  \end{tabular}
\end{table*}

\begin{table*}[t]
  \centering
  \caption{Scenario gist and most endorsed non-trust reasons (number of respondents).}
  \label{tab:scenario-notrust}
  \begin{tabular}{p{0.06\textwidth}p{0.32\textwidth}p{0.54\textwidth}}
    \toprule
    Scenario & Gist & Most frequently selected non-trust reasons (count) \\ \midrule

    A &
    Code-first, then reconstruct a higher-level model from LLM-generated code to understand and verify behavior. &
    The LLM lacks knowledge about the specific context (11);
    the LLM may oversimplify complex problems (9);
    explanations may sound convincing without being correct (8);
    the LLM does not communicate uncertainty or limitations (8);
    important details or edge cases may be overlooked (8). \\[0.6em]

    B &
    Use model reconstruction on an existing long-running code base to understand the system before continuing development with LLM-generated code. &
    Important details or edge cases may be overlooked (7);
    the LLM may generate misleading or fabricated information (6);
    explanations may sound convincing without being correct (6);
    the LLM does not communicate uncertainty or limitations (6);
    the LLM may oversimplify complex problems (6). \\[0.6em]

    C &
    Update an existing system’s model (based on current code and a new requirement) with an LLM, review the model, then generate code changes. &
    The LLM may generate misleading or fabricated information (6);
    explanations may sound convincing without being correct (6);
    the LLM may oversimplify complex problems (6);
    the LLM lacks knowledge about the specific context (5);
    the LLM cannot take responsibility for mistakes (5). \\[0.6em]

    D &
    Generate and review system models (architecture, data, behavior) with an LLM for a new system before deterministic code generation. &
    The LLM may generate misleading or fabricated information (7);
    the LLM does not communicate uncertainty or limitations (7);
    the LLM may oversimplify complex problems (7);
    explanations may sound convincing without being correct (5);
    the LLM lacks knowledge about the specific context (5). \\

    \bottomrule
  \end{tabular}
\end{table*}

\subsection{RQ3: Perceptions of Models and Vibe Modeling as Intermediate Abstractions}
Across all four scenarios, students consistently treated models as valuable intermediates between natural-language requirements and AI-generated code. The trust analyses show that participants repeatedly emphasized reasons such as detailed and precise results, transparent intermediate reasoning, and the ability to independently verify outcomes, all of which are properties more naturally associated with model-based views than with raw code.\
In scenarios where models are reconstructed from existing code (A, B), updated before code changes (C), or generated first for a new system (D), students tend to trust the workflow more when they can inspect and reason about a structured representation of the system before committing to implementation.

To probe perceptions of vibe modeling as an intermediate abstraction, we asked how useful students think it would be for understanding LLM generated code, validating LLM-generated code and for understanding code changes (1 = not useful, 101 = very useful).
7 students out of 17, answered these questions. 6 out of 7 rated above 60 for understanding LLM generated code with vibe models. For validation, 5 out of 7 students rated it above 50. 
For understanding code changes, 4 out of 7 rated it above 50.
Here, the perceived utility is slightly higher and more skewed towards strong usefulness, indicating that students particularly appreciate vibe modeling as a way to reason about how systems evolve over time.

Overall, these descriptive results align with our scenario-based findings: students see models—and specifically vibe models—as valuable intermediate artifacts that help them validate and understand AI-assisted changes, even though they also remain aware of the limitations and risks of relying on LLM-generated artifacts.
Open comments further underline that vibe models are seen as making AI-assisted development more interpretable and controllable, \enquote{\emph{I would trust it as I can see LLM create models, I think same as UML. So I can check and write code myself}}. Some students noted that learning to read and adjust code introduces additional effort \enquote{\emph{I am still learning, I guess having a model makes sense, sometime it's hard to read code and check.
With class diagram or such I can find the structure}}. Overall, the findings suggest that students first recognize the need for model-based intermediate abstractions to anchor trust and understanding, and then perceive vibe modeling as a promising way to obtain such models in AI-centered development workflows.
\enquote{\emph{LLM is not good doing consistent work like changing the model based on previous code and new changes, it might hallucinate. Other hand, Ben is checking the model before asking LLM to generate code from that model. This is better for trusting as I can catch LLM by the model if the latest code base is wrong}}\section{Discussion}
Our exploratory study set out to examine how students perceive model-based intermediates including vibe modeling in AI-assisted software development workflows, using four scenarios that varied in how and when models are introduced relative to LLM-generated code.

\subsection{Models as Trust Anchors in AI-Assisted Workflows}
Across scenarios, students systematically associated trust with properties that are more naturally expressed through models than through raw code: sufficient detail and precision, transparent intermediate reasoning steps, independent verifiability, and alignment with their own expectations or expertise.
In Scenario~A (code-first, then reconstructed model), trust reasons prominently featured “The LLM allows human oversight or correction”, “The LLM provides transparent reasoning or intermediate steps”, and “The results can be verified independently”, suggesting that students rely on both human control and in between abstraction layers to make sense of LLM-generated code.
Similarly, in Scenarios~C and~D—where models are updated before code changes or generated first for a new system, students highlighted detailed and precise results, transparent reasoning, and evidential support as key trust enablers, indicating that the presence of an explicit model layer is perceived as a way to “anchor” AI outputs in a structured explanation.

Non-trust reasons complement this picture: regardless of scenario, students repeatedly expressed concerns that the LLM might generate misleading or fabricated information, oversimplify complex problems, lack knowledge about the specific context, fail to communicate uncertainty, and overlook important details or edge cases.
Taken together, these patterns point to a tension that is well documented in broader trust-in-AI work: students appreciate the productivity and convenience of AI assistance, but distrust opaque, context-insensitive outputs that cannot be easily checked or traced~\cite{scharowski2025trust}
In our scenarios, models function as a means to resolve part of this tension by providing a more inspectable, discussable, and verifiable representation of the system than code alone.

\subsection{From General Models to Vibe Models}

Our scenarios intentionally combined “traditional” model generation (e.g., code-to-model reconstruction, model updates, model-first design) with the emerging idea of vibe modeling, where models themselves are obtained through conversational interaction with an LLM rather than handcrafted by developers~\cite{cabot2026vibe,nevskovic2026extending}. %
The data suggest that students first recognize the generic value of models as intermediate abstractions—independent of how they are obtained and then extend this appreciation to vibe models when they see them as supporting understanding, validation, and change reasoning.

Quantitative ratings of usefulness indicate that students perceive vibe modeling as highly useful for understanding LLM-generated code, validating results, and reasoning about code changes, with median scores above the mid-point of the scale for all three items.
Trust reasons in Scenarios~C and~D, which are closest to vibe-modeling workflows (model update before code change and model-first generation), emphasize transparent reasoning, detailed and precise results, and human oversight, aligning well with the conceptual promise of vibe modeling as “model-driven vibe coding” that places models—not code—as the primary locus of interaction and review.
At the same time, the persistent non-trust concerns about fabrication, oversimplification, and missing context remind us that simply introducing models is not sufficient; the quality and faithfulness of those models, and the mechanisms by which they are validated, are crucial for realizing the potential of vibe modeling in practice.

\subsection{Scenario-Based Insights: Where Models Help Most}
The four scenarios offer a progressive view of how models can be integrated into AI-assisted workflows:

\subsubsection{Scenario A (Code-first, then reconstruct model)} reflects a reactive use of models: students rely on reconstruction to make sense of opaque LLM-generated code after the fact.
        Trust is driven by oversight and transparent reasoning, but non-trust concerns emphasize missing context and oversimplification, suggesting that reversing from code to models is perceived as helpful yet limited when the underlying code may already embody problematic decisions.

\subsubsection{Scenario B (Code-to-model reconstruction for understanding)} foregrounds model reconstruction on an existing code base before continuing development.
        Here, trust reasons emphasize independent verification and alignment with the developer’s expertise, while non-trust reasons highlight the risk of fabricated information and overlooked details.
        This pattern suggests that reconstruction models are seen as useful for orientation, but students remain wary that AI-generated views might misrepresent legacy complexity or fail to surface critical edge cases.

\subsubsection{Scenario C (Model update before code change)} moves models “upstream” in the change process: students review an updated model before applying code changes.
        Trust in this scenario is linked to plausible, sufficiently detailed results, transparent reasoning, and human oversight, while non-trust reasons emphasize fabrication, oversimplification, and context gaps.
        This indicates that introducing a model checkpoint prior to code generation is perceived as a meaningful improvement, but only if the model itself is reliable and grounded in the actual system state.

\subsubsection{Scenario D (Model-first generation for a new system)} represents the cleanest realization of a model-based workflow: models are generated and iteratively refined before deterministic code generation.
        Students reward this scenario with trust reasons centered on detail, plausibility, transparent reasoning, evidential support, and verifiability, yet still articulate non-trust concerns about fabrication and limited uncertainty communication.
        This suggests that model-first workflows embody the direction students find most promising, but that the underlying LLM still needs guardrails to avoid superficially plausible but misleading designs.

Overall, the scenario-based analysis supports a design principle for AI-assisted development: placing models earlier in the workflow, and using them as explicit, reviewable checkpoints, tends to increase perceived trustworthiness and control, but does not by itself eliminate concerns about AI-generated artifacts that may be wrong, shallow, or context-insensitive.
Vibe modeling appears as a natural extension of this principle, provided it is coupled with robust validation and alignment mechanisms.

\section{Future Plans}

Building on these exploratory findings, we see several concrete directions for future work in trustworthy AI-assisted, model-based development.
We plan to extend this work along the following timeline:

\begin{itemize}
  \item \textbf{Short term (next 4 months).}
  Run larger surveys with more students at multiple universities, using more problem-oriented questions (e.g., concrete LLM usage and trust dilemmas) rather than purely scenario-based prompts.
  This will refine our measures and provide a more robust picture of student attitudes.

  \item \textbf{Medium term (4-10 months).}
  Conduct follow-up studies with professional developers, comparing student and practitioner responses on trust, non-trust, and perceived value of models and vibe modeling in real project contexts.

  \item \textbf{Long term (1+ years).}
    Design and evaluate prototype vibe-modeling tools and validation mechanisms for vibe models. The end goal is to combine data-driven insights on trust with deployed model-first workflows and training materials on model validation, AI risk assessment, and trust calibration.
\end{itemize}
\newpage
\bibliographystyle{ACM-Reference-Format}
\bibliography{refs}
\end{document}